\documentclass[fleqn,usenatbib]{mnras}

\usepackage{newtxtext,newtxmath}
\usepackage[T1]{fontenc}
\usepackage{ae,aecompl}

\usepackage{graphicx}	% Including figure files
\usepackage{amsmath}	% Advanced maths commands
\usepackage{amssymb}	% Extra maths symbols
\usepackage{natbib}
\usepackage[parfill]{parskip}
\usepackage{enumitem}
\usepackage[mathscr]{euscript}

\newcommand\tei{turbulent-entropic instability}
\newcommand\TEI{Turbulent-Entropic Instability}

\newcommand\uu{w}
\newcommand\UU{W}

\numberwithin{equation}{section}

\title[TEI]{A \TEI\ and the Fragmentation of Star-Forming Clouds}

\author[Keto, Field, Blackman]{
Eric Keto,$^{1}$\thanks{E-mail: eketo@cfa.harvard.edu (EK)}
George B. Field,$^{2}$
Eric G. Blackman$^{3}$
\\
$^{1}$Harvard University, Institute for Theory and Computation, 60 Garden St., Cambridge, MA 02138, USA\\
$^{2}$Harvard University, Department of Astronomy, 60 Garden St., Cambridge, MA 02138, USA\\
$^{3}$University of Rochester, Department of Physics and Astronomy, Rochester, NY,14627-0171, USA
}

\date{Accepted XXX. Received YYY; in original form ZZZ}

\pubyear{2019}

\begin{document}
\label{firstpage}
\pagerange{\pageref{firstpage}--\pageref{lastpage}}
\maketitle

% Abstract of the paper
\begin{abstract}
The kinetic energy of supersonic turbulence within interstellar clouds 
is subject to cooling by dissipation
in shocks and subsequent line radiation. The clouds are therefore susceptible to a condensation process controlled by
the specific entropy. 
In a form analogous to the thermodynamic entropy, the entropy for supersonic turbulence is proportional to the log of the
product of the mean turbulent velocity and the size scale.
We derive a dispersion relation for the growth of entropic instabilities
in a spherical self-gravitating cloud and find that there is a critical maximum dissipation time scale, about equal to the crossing time, 
that allows for fragmentation and subsequent star formation. 
However,  
the time scale for the loss of turbulent energy may be shorter or longer, for example with rapid thermal cooling or the injection of mechanical energy. 
Differences in the time scale for energy loss
in different star-forming regions may result in differences in the outcome, for
example, in the initial mass function.  
\vspace{1cm}
\end{abstract}
% Select between one and six entries from the list of approved keywords.
% Don't make up new ones.

%%%%%%%%%%%%%%%%%%%%%%%%%%%%%%%%%%%%%%%%%%%%%%%%%%

%%%%%%%%%%%%%%%%% BODY OF PAPER %%%%%%%%%%%%%%%%%%
\section{Introduction}

Star formation requires that smaller, self-gravitating regions dynamically separate from a larger interstellar cloud into individual centers of collapse.
Several hypotheses explain how this fragmentation process might happen.
The {gravitational cascade}  \citep{Hoyle1953} supposes that because gravitational contraction occurs on a free-fall timescale $\sim 1/\sqrt{G\rho}$ where $\rho$ is the local density,  higher density regions contract on a shorter timescale allowing a collapsing cloud to develop a hierarchy of smaller and denser collapsing subregions.   
The hierarchy is described as "gravitational turbulence" emphasizing the dominance of gravitational over hydrodynamic forces. More
recent developments on this idea are summarized in \citet{VS2019}.
Alternatively, the hypothesis of {gravoturbulent fragmentation} \citep{Elmegreen1993, Klessen2000} suggests that small regions of individual collapse 
form where gas between 
turbulent eddies  is
compressed beyond the local Jeans density.
The {thermal instabity}  \citep{Field1965} allows the possibility of a bi-stable interstellar medium (ISM) 
with a cold ($\sim$ 100 K), dense phase co-existing in pressure 
equilibrium with a hot (few 1000 K) rarefied phase.
On cooling, the hot phase can fragment into individual cold clouds that could gravitationally collapse to form stars
\citep{Hunter1966}. However, star formation generally occurs in molecular clouds and this process is not applicable to the fragmentation 
of the molecular phase itself. Nonetheless, the thermal instability may
play a role in the development of turbulence in molecular clouds formed from warmer atomic gas and thus in setting 
the initial conditions for fragmentation \citep{Heitsch2005}.

In this paper we discuss fragmentation by a different process. Similar to fragmentation by the thermal instability, 
we imagine a condensation process defined by entropy fluctuations. 
At the low temperatures (10 - 25 K) typical of molecular clouds, the internal energy is dominated by the kinetic energy of 
supersonic turbulence. For example, the energy associated with a turbulent velocity of 2 kms$^{-1}$ corresponds to 
the energy of a quiescent cloud at a temperature of 1100 K. 
Therefore the relevant cooling or loss of energy is
the dissipation of turbulent energy in shocks and subsequent line radiation. 
Accordingly, we define an entropy for turbulence analogous to the thermal entropy
but related to turbulent energy rather than thermal energy.

Entropy fluctuations are unstable in a thermal self-gravitating cloud
because of its negative heat capacity. The negative heat capacity of a star is
well understood. As a star loses total energy by radiation, the contraction increases 
the thermal velocities, equivalent to an increase in temperature, to maintain equilibrium
with the increase in the absolute value of the potential energy. Meanwhile, the entropy decreases
 with the decrease in total energy.
More generally, if part of a cloud loses energy, 
then within this contracting region, the temperature rises while the entropy decreases. 
Conversely, a part of the
cloud that gains energy, expands with a resulting decrease in temperature
and an increase in entropy. The temperature difference results in
the transfer of heat or energy from the warmer region to the cooler region
further enhancing the temperature difference owing to the negative
heat capacity resulting in a condensation
instability. 

In a turbulent cloud, there is similar instability substituting the kinetic energy of
the turbulent velocities for the temperature. However, conduction or the transfer of
energy from one part of the cloud to another is not required for instability. 
Both the contracting perturbation and the larger scale cloud are losing energy
through turbulent dissipation. However, the contracting perturbation loses energy at an 
increasingly faster rate, 
inversely proportional to the
turbulent crossing time.
It is easy to see that a decrease in the crossing time 
resulting from an increase in the mean turbulent velocity and a decrease in size
enhances the difference in the entropy between the
contracting perturbation and the rest of the cloud. 
The result is the
turbulent entropic instability.

The instability is best described by the entropy which continuously decreases
along with the total energy. In contrast, the relationship between the
mean turbulent velocity and the radius can be more complex.  
Dissipation continuously decreases the mean turbulent velocity
while contraction has the opposite effect.
The relationship then depends
on the ratio of the crossing time and the gravitational time. These are 
approximately the same in equilibrium but may not remain so as the instability
develops. Also, the loss rate of turbulent energy in a star-forming cloud
may be modified by other processes. 
For example, an input of kinetic energy
from star-formation feedback or shear may increase the time scale for energy loss by supplying
fresh turbulent energy. We find that if the loss time is longer than a 
critical value about equal to the crossing time or two times the gravitational time, the instability does not operate. 
In physical terms, sufficiently long loss times allow the cloud to erase perturbations dynamically 
before they have time to grow.

A necessary condition for fragmentation, defined as the development of dynamically distinct 
subregions, is that the larger scale contracts more slowly than
the smaller. This condition is satisfied by the \tei\ as well as by the gravitational instability. As \citet{Hoyle1953} points out,
fragmentation may be extended to multiple scales resulting in a hierarchy or cascade 
\footnote{The process of hierarchical fragmentation may be described as a transfer or
cascade of mass from larger to smaller scales. A description is given in \citet{Field2008} for an assumed relationship
between the mean turbulent velocity and the radius and a steady-state over the mass scales.}.
In the gravitational instability, the time scale is proportional to the inverse of the 
square root of the density while in the \tei\ the time scale is inversely proportional to the
crossing time. If the mass of the contracting region is considered a constant, the density and radius are of course linked.
The \tei\ can replace the gravitational instability as a driver of hierarchical fragmentation.

By way of comparison, the \tei\  is similar to fragmentation through the thermal instability in
that both are a condensation process. However, the \tei\ can only occur in a self-gravitating cloud and does not result in a phase change. 
The cooling rate in the \tei\ is related to the dynamics whereas the thermal instability depends on the shape of thermal cooling curve 
which is the sum of the atomic and molecular line cooling rates.
In both the \tei\ and thermal instability, the ratio of the dynamical time and the cooling time determines the dynamical outcome.

The complete outline of the paper is as follows.
\begin{description}
\item 
(Section \ref{ENTROPY}) There is a turbulent entropy $S \propto \ln (\sigma R)$ that is analogous to the thermodynamic entropy defined
by the first law, $dQ = TdS$ where $\sigma R$ is
the product of the turbulent velocity and
radius.
\item 
(Section \ref{DISSIPATION}) The rate of change of entropy is defined in terms of the turbulent dissipation rate. 
\item 
(Section \ref{TDVT})  
The equation governing the evolution of a spherical, self-gravitating, turbulent, molecular cloud  is derived from
a time-dependent form of the virial theorem. 
\item  
(Section \ref{DISPERSION}) Perturbations about equilibrium are described by damped oscillations with a frequency related to the 
gravitational time and the damping time related to the turbulent dissipation time. 
\item 
(Section \ref{FRAGMENTATION}) Fragmentation is defined in terms of a time-dependent Jeans mass.
\item 
(Section \ref{EVOLUTION})
There is a critical time scale for the decay of turbulent energy that allows for fragmentation. In terms of the crossing-time, $t_D/t_X < 0.7$ or in terms of the gravitational free-fall time,
$t_D/t_F < 2.1$ to allow for fragmentation. 
\item 
(Section \ref{DISCUSSION}) Differences in the effective loss rate of turbulent energy may explain different outcomes in different star-forming regions.
\end{description}

\section{Equations}
\subsection{Entropy}\label{ENTROPY}
From the first law of thermodynamics,
\begin{equation}
\frac {dQ} {dt} = \frac {dE} {dt} + P \frac {dV} {dt} .
\end{equation}
If the energies and volume are expressed per unit mass, then the energy and pressure in
a turbulent cloud with one-dimensional velocity dispersion $\sigma$, are,
\begin{equation}\label{kedef1}
E =K = \frac {3}{2} \sigma^2
\end{equation}
\begin{equation}\label{prdef}
P = \frac{1}{V}\sigma^2,
\end{equation}
where $K$ is the turbulent kinetic energy.
Therefore, the heat transfer per unit mass in the turbulent cloud is,
\begin{equation}\nonumber
\frac {dQ}{dt} =  \frac{3}{2} \frac{d\sigma^2}{dt} + \sigma^2 \frac{3}{R} \frac{dR}{dt} 
=  K \bigg({\sigma^{-2}}\frac{d\sigma^2}{dt} + \frac{2}{R} \frac{dR}{dt} \bigg),
\end{equation}
and
\begin{equation}\label{dqdt}
\frac {dQ}{dt} 
= 2K \frac{d}{dt} \ln (\sigma R) .
\end{equation}
In this form, the quantity $\ln(\sigma R)$ is analogous to the thermodynamic or "heat engine" entropy in the equation, $dQ/dt = TdS/dt$ with the 
kinetic energy analogous to the thermal temperature, $T$. 

The thermodynamic entropy in the first law describes the macrostate of the system.  A comparison with the 
distribution of microstates in 
Boltzmann's entropy explains why the turbulent form of the entropy is related to the product, $\sigma R$. 
In Boltzmann's entropy, $S=k\ln \mathcal{W}$, with the number of microstates, $\mathcal{W} \propto V E^{3/2}$.  
Because only changes in entropy are important, we can substitute $R^3$ for $V$ without loss of generality. 
We can also replace the energy per unit mass, $E$, 
by the turbulent velocity dispersion, $\sigma^2$. Then $\mathcal{W}\propto \sigma^3 R^3$ is proportional to the volume of 
phase space or equivalently the 
number of microstates. With these substitutions,
Boltzmann's entropy for turbulence  becomes,
\begin{equation}
S = k \ln( \sigma R) .
\end{equation}

\subsection{Rate of change of the turbulent entropy as a function of the rate of turbulent dissipation}\label{DISSIPATION}

To determine the evolution
of a cloud subject to turbulent dissipation, we derive the change in the turbulent entropy as a function of the
dissipation rate. 
The change in $\UU = \sigma R$ is equivalent to the change in the turbulent entropy with a non-linear rescaling, 
The rate of change of the kinetic energy is equal to the rate of turbulent dissipation,
\begin{equation}\label{dkdt}
\frac {dK}{dt} = \frac{dQ}{dt} = -\frac {K}{t_D}.
\end{equation}
The dissipation rate of compressible or incompressible turbulence is on the order of the crossing time \citep{Kolmogorov1941,Gammie1996}.
If the effective dissipation time scale, $t_D$, is some multiple, $\gamma$, of the crossing time, $t_X$,
\begin{equation}\label{tddef}
t_D = \gamma t_X =  \gamma \frac{R}{\sigma},
\end{equation}
then from equation \ref{dqdt} and \ref{tddef},
\begin{equation}
\frac {dQ}{dt} = 2K \frac{1} {\sigma R} \frac {d(\sigma R)}{dt} = 2K \frac {1}{\UU} \frac {d\UU}{dt}
\end{equation}
and with equation \ref{tddef},  the rate of change of $\UU$ itself is,
\begin{equation}\label{dudt}
\frac {d\UU}{dt} = -\frac{1}{2} \frac{\UU}{t_D} = -\frac{1}{2\gamma} \frac{\UU ^2 }{R^2} .
\end{equation}
The rate of heat loss $dQ/dt$ in equation \ref{dkdt} may be modified by other processes.
The decay rate of supersonic turbulence
may be faster than the crossing time if the gas temperature is also decreasing and keeping the Mach
number high even as the turbulent velocities decay \citep{Pavlovski2002}.  Alternatively, the addition of turbulent
energy from external or internal sources such as shear or star-formation feedback may slow the loss rate. 
In this case,
$dK/dt$ should be thought of as an effective dissipation rate or decay rate that is specified by the
factor $\gamma$ in equation \ref{tddef}. 
For example, if
\begin{equation}
\frac{dL}{dt} = \frac{L}{t_L}
\end{equation}
is the rate of increase in turbulent kinetic energy from an outside source of mechanical energy, then
$\gamma t_X$ may be defined as the effective time scale for the rate of change of the
turbulent kinetic energy due to both turbulent dissipation and fresh mechanical energy.
\begin{equation}
\gamma t_X = K\bigg(\frac{t_L t_D}{t_DL - t_LK}\bigg)
\end{equation}
At present, our knowledge of interstellar turbulence is insufficient to
specify $L/t_L$, and we leave this uncertainty expressed in the factor $\gamma$.

\subsection{The time-dependent virial theorem as a function of turbulent entropy}\label{TDVT}

Virial equilibrium implies that the gravitational and turbulent
kinetic energies are roughly in equipartition. 
The evolution of this relationship is described through a time-dependent form of the virial theorem.
The applicability assumes that the gravitational potential energy
can be converted into virialized kinetic energy on a dynamical time scale, for example by the transport of
angular momentum from smaller to larger scales through gravitational torques \citep{Henriksen1984}.
Following \citet{ChandraFermi1953} 
we interpret the virial theorem in a Langrangian sense integrating over a fixed mass.
An
external pressure may be incorporated either through a decrease in the kinetic energy or an increase in the gravitational force
inside the surface. Similarly, we assume that if there is
significant energy from a turbulent magnetic field, it is in equipartition with the turbulent kinetic energy and can be included with this term.  
In this case, the virial theorem per unit mass is,
\begin{equation}
\frac {1}{2} \frac{d^2I}{dt^2} = 2K + \Omega.
\end{equation}
Here
\begin{equation}\label{betadef}
I = \frac{1}{M} \int _0^M r^{\prime 2}dm = \beta R^2
\end{equation}
is the moment of inertia per unit mass, and $R$ is the radius with enclosed mass $M$. The kinetic energy, $K$, is the same as
defined earlier (equation \ref{kedef1}) if $\sigma$ is a function of time only,
\begin{equation}\label{kedef2}
K = \frac{1}{M} \frac{3}{2} \int _0^M \sigma^2 dm = \frac{3}{2} \sigma^2 .
\end{equation}
The gravitational potential energy is 
\begin{equation}\label{pedef}
\Omega = -\int _0^M r^\prime\frac{d \Phi}{dr^\prime}dm =-\frac{\Gamma G M}{R} .
\end{equation}
The factors, $\beta$ and $\Gamma$ are of order unity dependent on the internal
distribution of mass
\footnote{Appendix \ref{APPENDIXA} 
has calculations for the numerical value of $\beta$ 
for a power law density profile and for a hydrostatic density profile.}.
In terms of the variable pairs $R$, $\sigma$ or $R$, $W$ we have,
\begin{equation}
\frac {d^2(R^2)}{dt^2} = \frac{6}{\beta} \bigg( \sigma^2 - \frac{\Gamma G  M^2}{3R} \bigg) =  
\frac{6}{\beta} \bigg( \frac{\UU^2}{R^2} - \frac{\Gamma G  M^2}{3R} \bigg) .
\end{equation}
This equation can be written in non-dimensional form with scaling factors, $R_0$ and $\sigma_0$ for the 
radius and turbulent velocity dispersion respectively. Then $r=R/R_0$ and $s=\sigma/\sigma_0$, and these imply,
$\tau = t(\sigma_0/R_0)$ and $\uu=\UU/(R_0\sigma_0)$. 
Any combination of $\sigma_0$ and $R_0$ that satisfies virial equilibrium,
\begin{equation}\label{NDconstant}
\sigma_0^2 R_0 = \frac{\Gamma GM}{3} ,
\end{equation}
results in the non-dimensional equation,
\begin{equation}\label{ndvirial}
\frac{d^2r^2}{d\tau^2} = \frac{6}{\beta}\bigg(\frac{\uu^2}{r^2} - \frac{1}{r}\bigg).
\end{equation}

\subsection{Dispersion relation}\label{DISPERSION}

\begin{figure}
\includegraphics[width=3.25in]{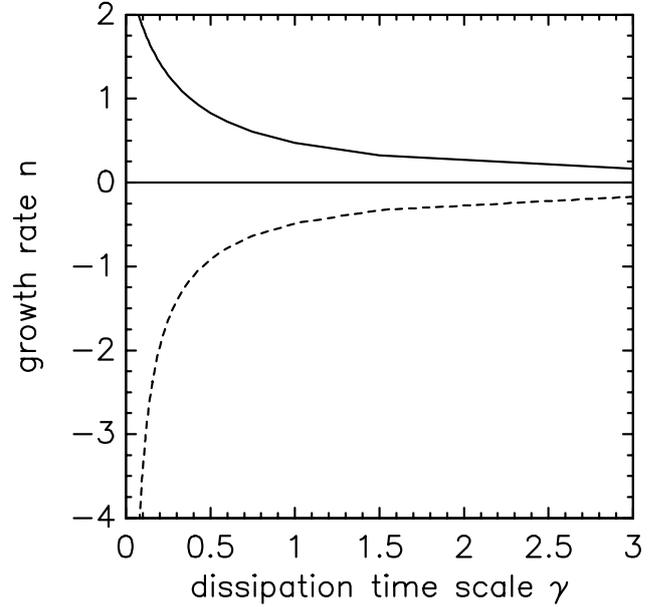}
\caption{
Growth rates of perturbations as functions of the dissipation time scale. 
The real parts of the 3 solutions of equation \ref{cubic} are plotted as functions of the dissipation time scale $\gamma$ in units of the crossing time.
The purely real solution is shown as a solid line and the real parts of the imaginary solutions, equal to each other, as a dashed line.
}
\label{gammaplot}
\end{figure}

The response of the cloud to small perturbations defines its stability.
With perturbations of the form, $r = 1+\delta_r \exp{(n\tau)}$ and $w = 1+\delta_w \exp{(n\tau)}$, then to first order,
equation \ref{ndvirial} becomes,
\begin{equation}\label{perturbed}
\frac{\beta}{6} \frac{d^2}{d\tau^2} ( 1 + 2 \delta_r \exp{(n\tau)}) =  
\frac{1 + 2\delta_w \exp{(n\tau)}}{1+ 2\delta_r \exp{(n\tau)}} - \frac{1}{1+\delta_r \exp{(n\tau)}} .
\end{equation}
The zero order terms cancel, leaving an equation with solutions proportional to $\exp(n\tau)$. Since $d^2/d\tau^2 \rightarrow n^2$,
equation \ref{perturbed} becomes,
\begin{equation}\label{field1}
\bigg( \frac{\beta}{3} n^2 - 1 + 2  \bigg) \delta_r = 2 \delta_w .
\end{equation}
In a similar way, we can include perturbations in equation \ref{dudt} for the rate of change of the turbulent entropy (rescaled as $w$).
First rewrite equation \ref{dudt} in non-dimensional form,
\begin{equation}\label{dudtnd}
\frac{dw}{d\tau} = -\frac{1}{2\gamma} \frac{w^2}{r^2}.
\end{equation}
The perturbed equation is,
\begin{equation}\label{field2}
\bigg(n + \frac{1}{\gamma}  \bigg) \delta_w = \frac{1}{\gamma}  \delta_r.
\end{equation}
Combining equations \ref{field1} and \ref{field2},
\begin{equation}\label{cubic}
\frac{\beta}{3} n^2 \bigg(n + \frac{1}{\gamma} \bigg) + n - \frac{1}{\gamma} = 0.
\end{equation}
This cubic equation for the growth rate has three solutions, one real, and two that are complex conjugates of each other.
The real solution indicates collapse while the imaginary solutions indicate damped oscillations.
The time scale for collapse is related to the change in $\uu$,  equation \ref{dudtnd},
and thus to the crossing time
while the time scale for the oscillations is the gravitational time scale.
Figure \ref{gammaplot} plots the growth rate, $n$, as a function of the dissipation time. 
In these examples $\beta = 0.4$, appropriate for a self-gravitating cloud (appendix \ref{APPENDIXA}). 
Comparing initial clouds of different initial crossing times, the growth rate $n$ is 
larger (faster) for shorter crossing times.
This property allows for hierarchical fragmentation because 
subregions can dynamically separate from a larger region.
%/Users/dog/papers/field/Sept-18/gamma_plot/gammaplot.pro
%
\subsection{Fragmentation}\label{FRAGMENTATION}
A self-gravitating cloud or region may fragment  into smaller self-gravitating regions if its mass
exceeds the Jeans mass. If we define the Jeans mass as in \citet{Spitzer1978}, page 283,
\begin{equation}
M_J = \frac {\pi^{3/2}\sigma^3} {G^{3/2} \bar\rho^{1/2}}
\end{equation}
then,
\begin{equation}
\bigg(  \frac {M_J}{M} \bigg) \approx \bigg( \frac{ M_V} {M} \bigg) ^{3/2}
\end{equation}
where $M_V$ is the virial mass,
\begin{equation}
M_V = \frac {3\sigma^2 R}{\Gamma G}
\end{equation}
defined from equations \ref{kedef2} and \ref{pedef}, and $\bar\rho$ is the mean density. The equality is
nearly exact if $\Gamma = 3/5$ as for a spherical cloud of uniform density. 

From an initial state of equilibrium, $M = M_V  \approx M_J$, the virial mass and the Jeans mass will both decrease because of
the decay of turbulent kinetic energy and contraction. However, the Jeans mass decreases faster by a power of $3/2$.
Suppose the cloud might fragment into two regions if $M_J / M \leq 2$.  In our non-dimensional variables, according to equation \ref{NDconstant}
this condition is,
\begin{equation}
\bigg( \frac{M_J} {M} \bigg)^{-1} = \bigg( \frac{M_V} {M} \bigg) ^{-3/2} = (s^2 r)^{-3/2} = \bigg(\frac {r} {\uu^2} \bigg)^{3/2}  \geq 2.
\end{equation}
Alternatively, $M/M_J = N$, the number of potential fragments.

\subsection{Non-linear Evolution}\label{EVOLUTION}

\begin{figure*}
\includegraphics[width=6.25in]{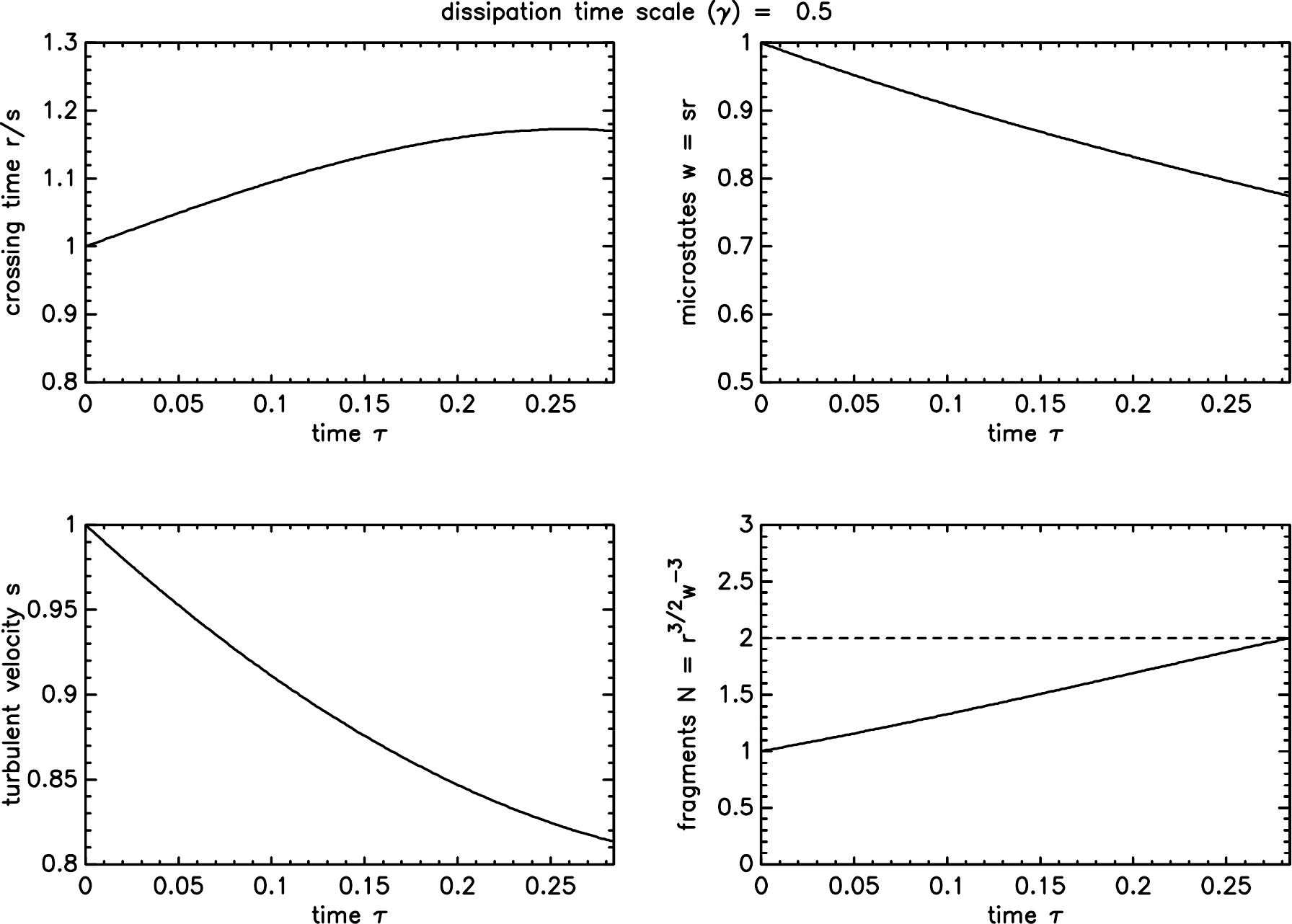}
%\vskip -2in
\caption{
Evolution of a cloud with a turbulent dissipation time scale $\gamma=0.5$ in equation \ref{tddef} or half the initial crossing time.
The evolution effectively ends when $N=2$, and the cloud fragments.
}
\label{evolution-0p5}
\end{figure*}

\begin{figure*}
\includegraphics[width=6.25in]{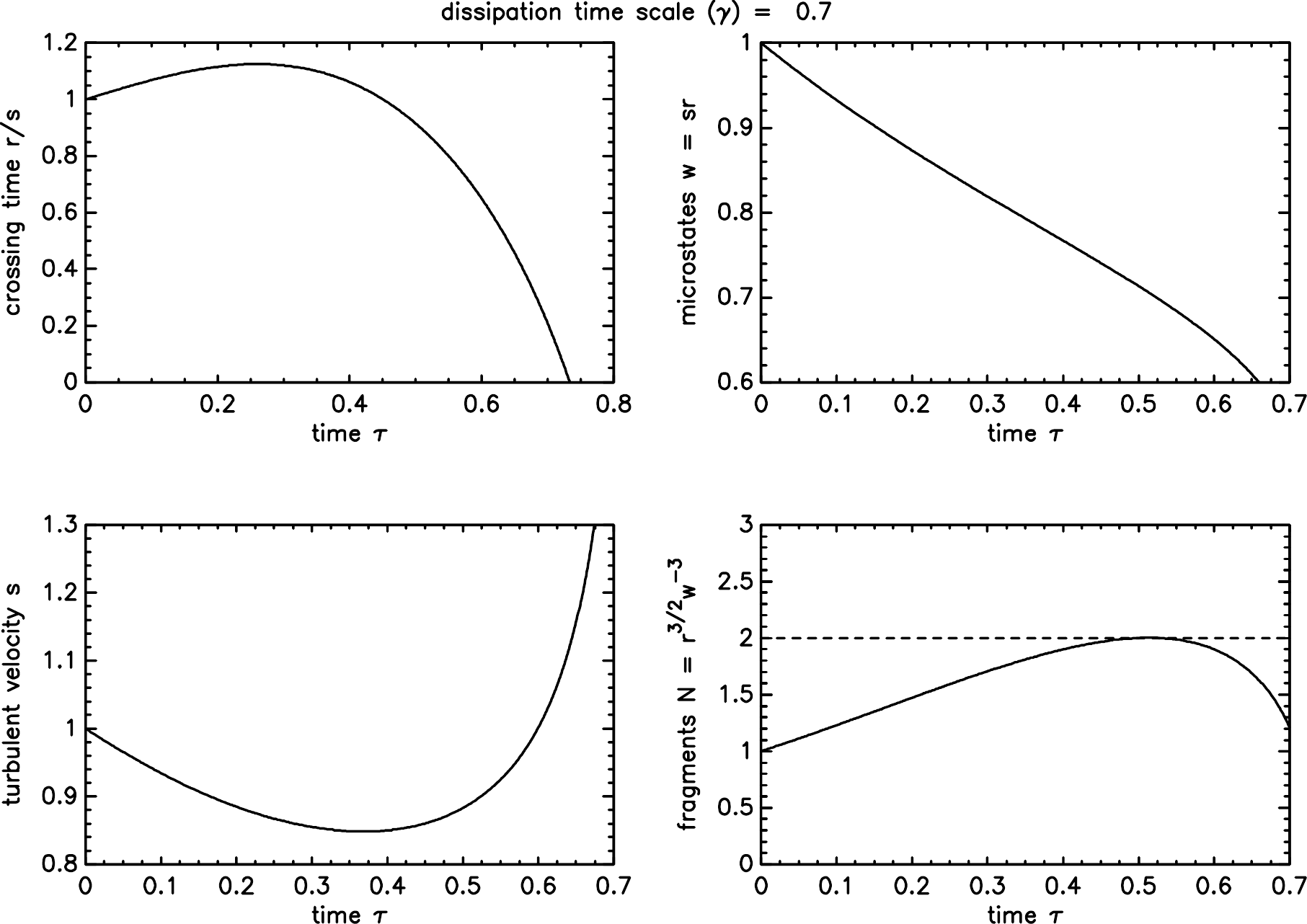}
%\vskip -2in
\caption{
Evolution of a cloud with a turbulent dissipation time scale $\gamma=0.7$ in equation \ref{tddef} or the initial crossing time.
This is the critical value that allows for fragmentation.
}
\label{evolution-1p0}
\end{figure*}

\begin{figure*}
\includegraphics[width=6.25in]{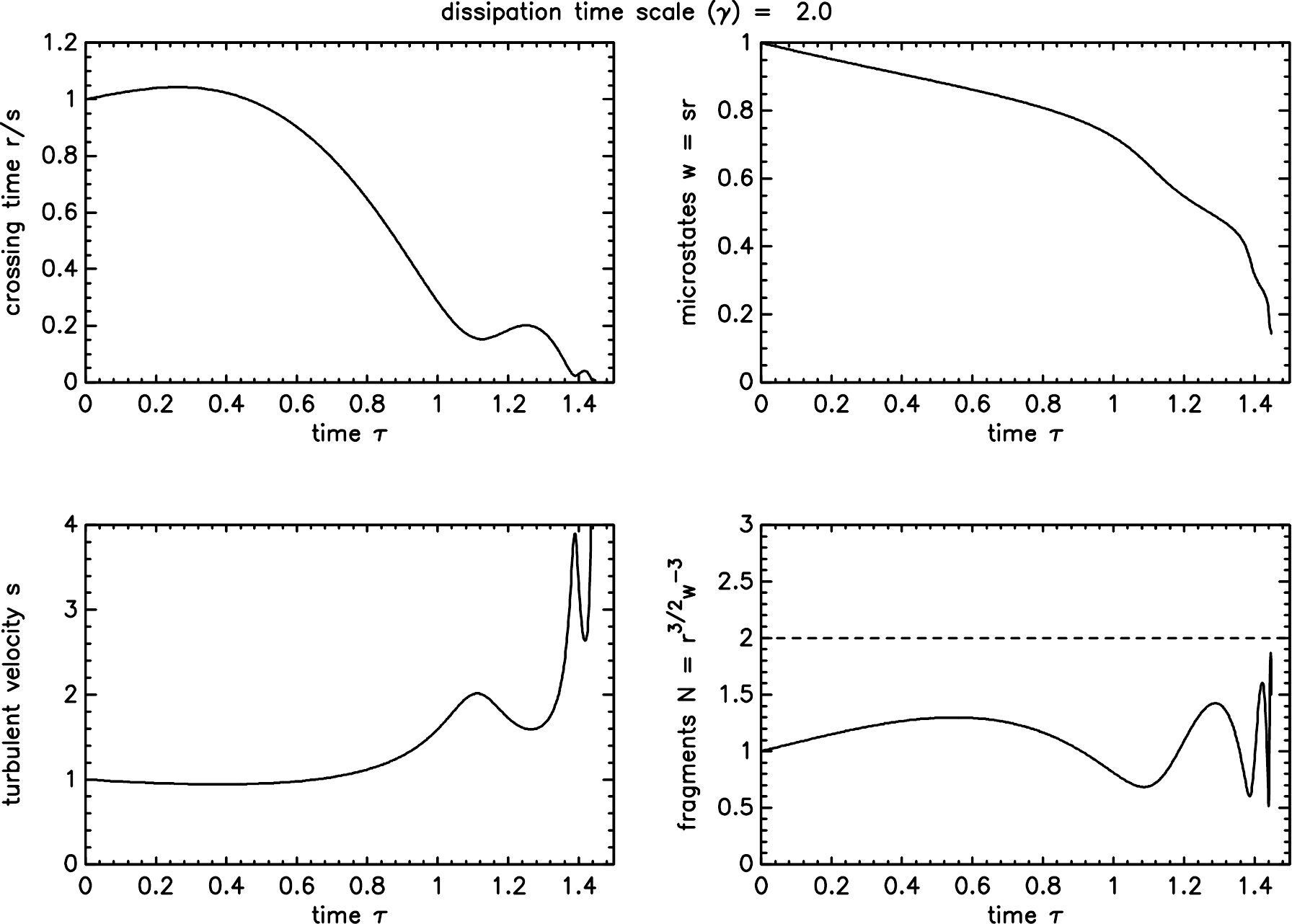}
%\vskip -2in
\caption{
Evolution of a cloud with a turbulent dissipation time scale $\gamma=2.0$ in equation \ref{tddef} or twice the initial crossing time.
}
\label{evolution-2p0}
\end{figure*}

The virialization or equipartition of the turbulent kinetic
and potential energies maintains a relationship between the radius of the cloud and the  turbulent velocities
described by the time-dependent virial equation  \ref{ndvirial}.  
We can numerically solve for this evolution more easily by converting this second order differential equation into three first order
differential equations. To do so, define the time evolution of the radius of the cloud as,
\begin{equation}\label{velocity}
\frac{dr}{d\tau} = v.
\end{equation}
So that,
\begin{equation*}
\frac{d^2r^2}{d\tau^2} = \frac{d}{d\tau}\bigg(2r\frac{dr}{d\tau}\bigg) = 2v^2 + 2r\frac{dv}{d\tau}.
\end{equation*}
Then, as a first-order differential equation, the time-dependent virial equation \ref{ndvirial} is,
\begin{equation}\label{acceleration}
\frac{dv}{d\tau} = \frac{1}{2r}\bigg(-2v^2 + \frac{6}{\beta}\bigg[\frac{\uu^2}{r^2} - \frac{1}{r}\bigg]\bigg).
\end{equation}
The three first-order differential equations to be solved are then \ref{velocity}, \ref{acceleration}, and \ref{dudtnd}.
The numerical solution shows that there is critical value for $\gamma < 0.7$ that allows fragmentation,
defined as $N \geq 2$. 
In terms of the crossing time, fragmentation is possible if the turbulent dissipation time scale is just shorter than the crossing time,
\begin{equation}
\frac {t_D}{t_X} < 0.7 .
\end{equation}
Appendix \ref{APPENDIXB} shows that the non-dimensional gravitational free-fall time is $\tau_F = \tau_X /3$. 
Therefore in terms of the gravitational free-fall time, fragmentation is possible if 
\begin{equation}
\frac{t_D} {t_F} < 2.1 .
\end{equation}
The exact numerical value, 0.7 or 2.1,
may not be significant given the spherical geometry assumed for the calculation. The critical value also depends on the 
density profile through the value of $\beta$ (appendix \ref{APPENDIXA}). The calculations shown in
figures \ref{evolution-0p5} through \ref{evolution-2p0} with $\beta =0.4$ are appropriate for an $r^{-2}$ power-law density profile
characteristic of self-gravitating clouds (Bodenheimer \& Sweigart 1968). 
These figures show the evolution for three different time scales $\gamma = 0.5, 0.7, 2.0$.
Figure \ref{evolution-0p5}  shows that  $\gamma=0.5$ allows rapid fragmentation with the subregions inheriting lower
turbulent velocities. We imagine that this would lead hierarchically to a large number of small fragments. The
fragmentation cascade will stop when the clouds are small enough to be supported by thermal energy and subsonic turbulent energy. 
Figure \ref{evolution-1p0} shows the evolution with the critical value of  $\gamma=0.7$ which is largest value that allows
fragmentation into at least two subregions. 
Since the time before fragmentation is longer than the previous example with $\gamma=0.5$, the cloud contracts
further allowing the negative heat capacity of the self-gravitating cloud to begin to increase the turbulent velocities from a minimum
before
fragmentation.
Figure \ref{evolution-2p0} with $\gamma = 2$ shows that a long dissipation time scale
prevents fragmentation as the cloud has time to dynamically relax. In this case the entire cloud eventually collapses with
some oscillations. However, if the turbulent energy is continuously
resupplied so that the dissipation time scale is infinite, the cloud will neither collapse nor fragment.

%/Users/dog/papers/field/Sept-18/october/octoberplot.pro
%
%/Users/dog/papers/field/Sept-18/october/octoberplot.pro
%
%/Users/dog/papers/field/Sept-18/october/octoberplot.pro

\bigskip\bigskip
\section{DISCUSSION}\label{DISCUSSION}
\bigskip

\subsection{Bimodal star formation}

Our analysis of the the \tei\ shows that the evolution of a cloud depends on the time scale for the loss of turbulent energy. With a shorter time scale, a cloud
may fragment multiple times with each episode resulting in lower turbulent velocities (figure \ref{evolution-0p5}) and smaller masses. 
With a longer dissipation time scale, a cloud may fragment more slowly resulting in fewer fragments or eventually collapse with its original mass. Thus the final
mass of a star-forming fragment may depend on the time scale for turbulent dissipation. In the next two sections we compare different outcomes in
high-mass and low-mass star forming regions.

\subsection{Observations}

\subsubsection{The Galactic Center clouds}

\begin{figure}
\includegraphics[width=3.5in]{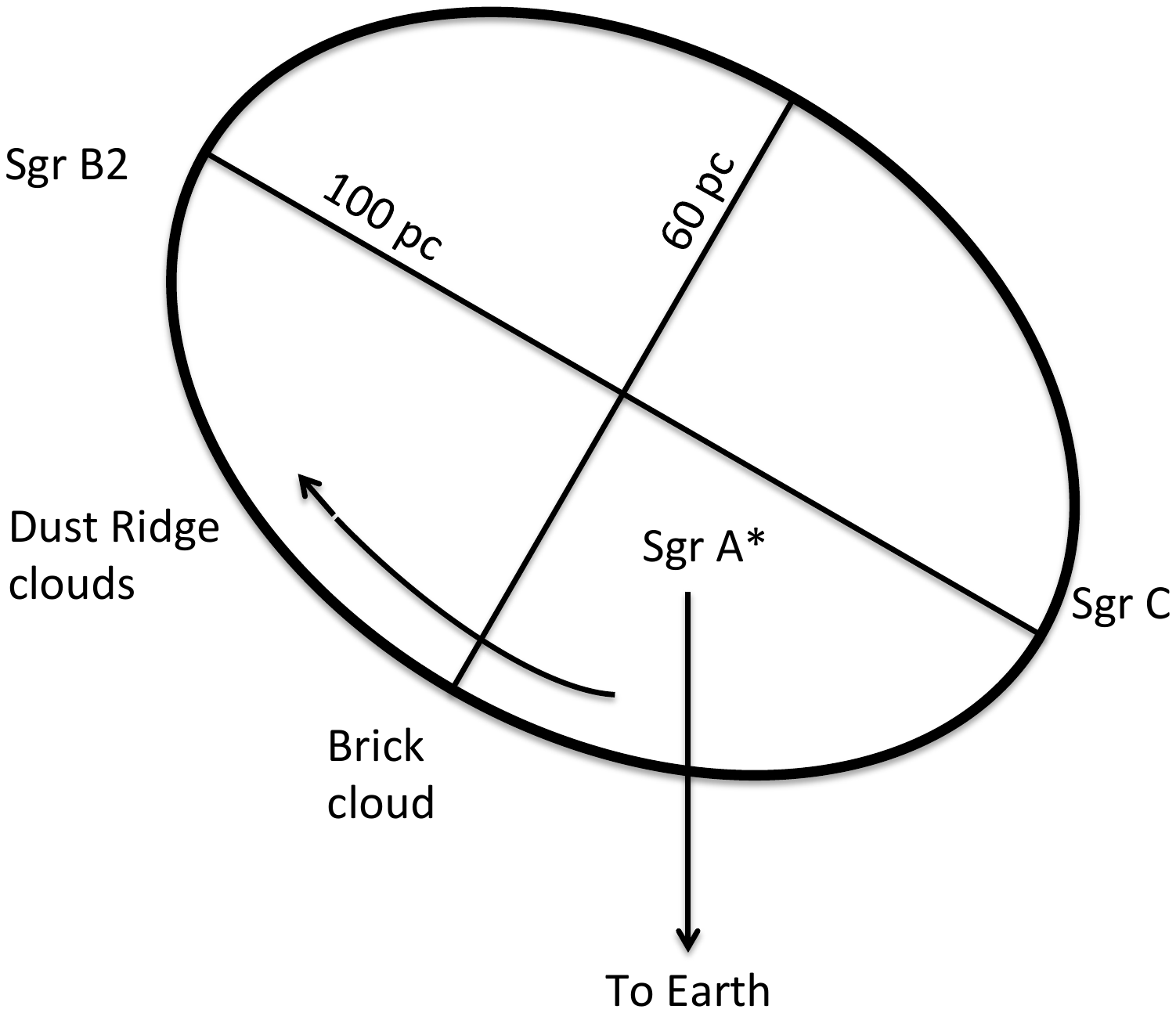}
%\vskip -2in
\caption{
Schematic of the clouds in elliptical orbit in the Galactic center adapted from \citet{Kruijssen2014}.
The view is of the plane of the Galaxy from above. (\citet{Kruijssen2015} suggest that the orbit is
not exactly elliptical in that the path from SgrC may not close back to the location of the Brick.)
}
\label{schematic}
\end{figure}

The clouds in elliptical orbit (60 by 100 pc) around the central black hole SgrA*
\citep{Molinari2011}
are of interest in star-formation studies because they defy "universal laws" of star
formation such as the Schmidt-Kennicutt relationship between the surface density and the star formation rate 
(\citet{Schmidt1959}; \citet{Kennicutt1998})
or the density threshold for star formation \citep{Lada2010}. In particular, the observed star formation rates for 
several of these clouds are an order of magnitude below those predicted by these relationships between the density and the rate
\citep{Longmore2013a}.
A case in point is the cloud 
GCM-0.02-0.07, (M$\sim10^5$ M$_\odot$) 
\citep{Johnston2014} nicknamed the Brick because of its high optical depth in the infrared and lack of
internal infrared emission from star formation. Only  3-4\% of the mass of this cloud is in dense star-forming regions often 
called high-mass cores.
The elliptical ring also includes the massive star-forming region Sgr B2 with 
some of the highest rates of star formation in the galaxy as well as the region Sgr C rich with HII regions 
indicating earlier active star formation. Figure \ref{schematic} shows the locations of the clouds on their elliptical orbit.

The clouds on the ring orbit in the direction from the Brick toward Sgr B2.
The Brick and other clouds with low star-formation rates are near the pericenter, 45 pc from the super-massive black hole
Sgr A$^*$ \citep{Johnston2014} 
while Sgr B2 is at the apocenter, twice as distant at 100 pc. \citet{Longmore2013b} suggest that the Brick and nearby clouds
are currently undergoing tidal compression which will result in intense
star-formation by the time the clouds arrive at the location of Sgr B2.  
The mini-starburst Sgr B2 would then be the result of compression in its earlier passage through the pericenter
of the orbit. Observations indicate that the clouds on the orbit between the Brick and Sgr B2, the Dust Ridge clouds, 
may be just beginning to fragment \citep{Walker2018}. 

Alternatively, fragmentation and ultimately the star formation rate may be controlled by the \tei .
If orbital shear slows the effective turbulent dissipation rate
by supplying fresh turbulent energy,
then the difference in orbital shear along the elliptical orbit from pericenter to apocenter
may prevent star formation near the pericenter and allow it at the apocenter. Following \citet{Kruijssen2014}, we derive a 
rotation curve, $v_{circ} = 23.5 r^{0.38}$ kms$^{-1}$ 
from the mass distribution measured around the Galactic center by \citet{Launhardt2002}.
At the radial distance from the Brick to the Galactic center, 45 pc, 
the orbital shear across a 10 pc diameter cloud is $\Delta v = 8.4$ kms$^{-1}$ while the one-dimensional
turbulent velocity dispersion in the Brick is 4.3 kms$^{-1}$ \citep{Johnston2014}. At the radial distance of SgrB2, 100 pc,
the shear is $3.4$ kms$^{-1}$ across 10 pc while the turbulent velocity dispersion in SgrB2 is 10.9 kms$^{-1}$ \citep{Henshaw2016}.
The higher ratio of shear velocity to turbulent velocity in the Brick may be suppressing fragmentation with respect to SgrB2. 

\subsubsection{Low-mass star formation in Taurus}

In contrast with the highly turbulent environment in the Galactic center, the Taurus star-forming region  \citep{Goldsmith2008} is relatively quiet with
CO line widths $ \leq 2$ kms$^{-1}$.
The whole region is less massive (M$\sim 2.4\times 10^4$ M$_\odot$) than some single
starless clouds in the CMZ, but nonetheless
characterized by rapid star formation. There are on the order of 600 clouds of a few M$_\odot$, often called low-mass cores, 
and 200 low-mass stars $\leq 2$ M$_\odot$, 
and there are no stars of greater mass. The region is off the Galactic plane at a latitude of 25$^\circ$ and at Galactic radius of $\sim 8$ kpc, 
not strongly affected by shear due
to Galactic rotation. Magnetic fields are ordered at least in the lower optical depth gas where the polarization of background 
starlight can be measured. In summary, there appear to be few external sources of turbulent energy. 
The only obvious source of fresh turbulent energy is from star formation feedback in the form of bipolar outflows, 
which of course begin after fragmentation and star formation.

Of interest to theories of fragmentation is a subset of the low-mass cores without stars, the starless cores, which may be the
sites of future star formation. The starless cores appear to be in quasi-hydrostatic
equilibrium supported against their self-gravity by a combination of thermal pressure and subsonic turbulence as deduced by
narrow (near thermal) line widths seen in high density molecular gas tracers 
such as NH$_3$ \citep{Myers1983}.
The cores are clustered on scales of 4 or 8 pc. The velocity
dispersion of the cores is a power law function of their separation with a mean value of 1.2 kms$^{-1}$ at a core separation $\le 10$ pc 
\citep{Qian2012}. If we suppose the cluster scale to represent the scale of a cloud prior to fragmentation, and the core dispersion
velocity to represent the initial turbulent velocity dispersion, then the crossing time of the initial cloud would be
6 pc / 1.2 kms$^{-1}$ or 5 Myr. Smaller regions of the size of the cores themselves, 0.3 pc, with crossing times at least an order
of magnitude less, could  fragment out of the cluster-sized cloud. 

As noted in section \ref{DISSIPATION}, the turbulent dissipation rate may be enhanced in a cloud whose average thermal temperature is cooling.
The relatively low overall column density and optical depth of a low-mass star forming such as Taurus may have allowed efficient radiative cooling
if the gas were initially warmer than its steady state temperature. The enhanced turbulent dissipation rate would then have allowed rapid fragmentation
resulting in numerous small clouds.

\section{Conclusions}

We derived the energy equation for turbulence in a self-gravitating cloud including
the contribution from PdV work from contraction as a result of turbulent dissipation. 
This equation suggests a form of the entropy for turbulent gas with the quantity, $W=\sigma R$, the turbulent velocity dispersion and the length scale,
analogous to the number of microstates in the definition of the thermal entropy, $S = k\log \mathcal{W}$.
A dispersion relation based on a time-dependent form
of the virial equation allows both an unstable mode on the crossing time, $R/\sigma$, and oscillating modes on the gravitational time. 
Following the non-linear evolution 
of a self-gravitating cloud with the virial equation and the energy equation, 
we find that clouds are unstable to hierarchical fragmentation if the dissipation time is on the order of
the crossing time or less. 
If the effective dissipation time scale is longer, the clouds do not fragment. 
Differences in the dissipation time scale in regions where fragmentation is controlled by the  \tei\ may explain differences in the outcome of star-formation in 
quiescent regions such as Taurus and strongly sheared regions such
as the Central Molecular Zone around the Galactic center and perhaps provide an explanation for bimodal star formation.

%\section*{Acknowledgements}

%The Acknowledgements section is not numbered. Here you can thank helpful
%colleagues, acknowledge funding agencies, telescopes and facilities used etc.

%%%%%%%%%%%%%%%%%%%%%%%%%%%%%%%%%%%%%%%%%%%%%%%%%%

%%%%%%%%%%%%%%%%%%%% REFERENCES %%%%%%%%%%%%%%%%%%

% The best way to enter references is to use BibTeX:

\bibliographystyle{mnras}
\bibliography{bibliography} % if your bibtex file is called example.bib

%%%%%%%%%%%%%%%%%%%%%%%%%%%%%%%%%%%%%%%%%%%%%%%%%%

%%%%%%%%%%%%%%%%% APPENDICES %%%%%%%%%%%%%%%%%%%%%

\appendix
\section{Calculation of moment of inertia}\label{APPENDIXA}

\begin{figure}
\includegraphics[width=3.25in]{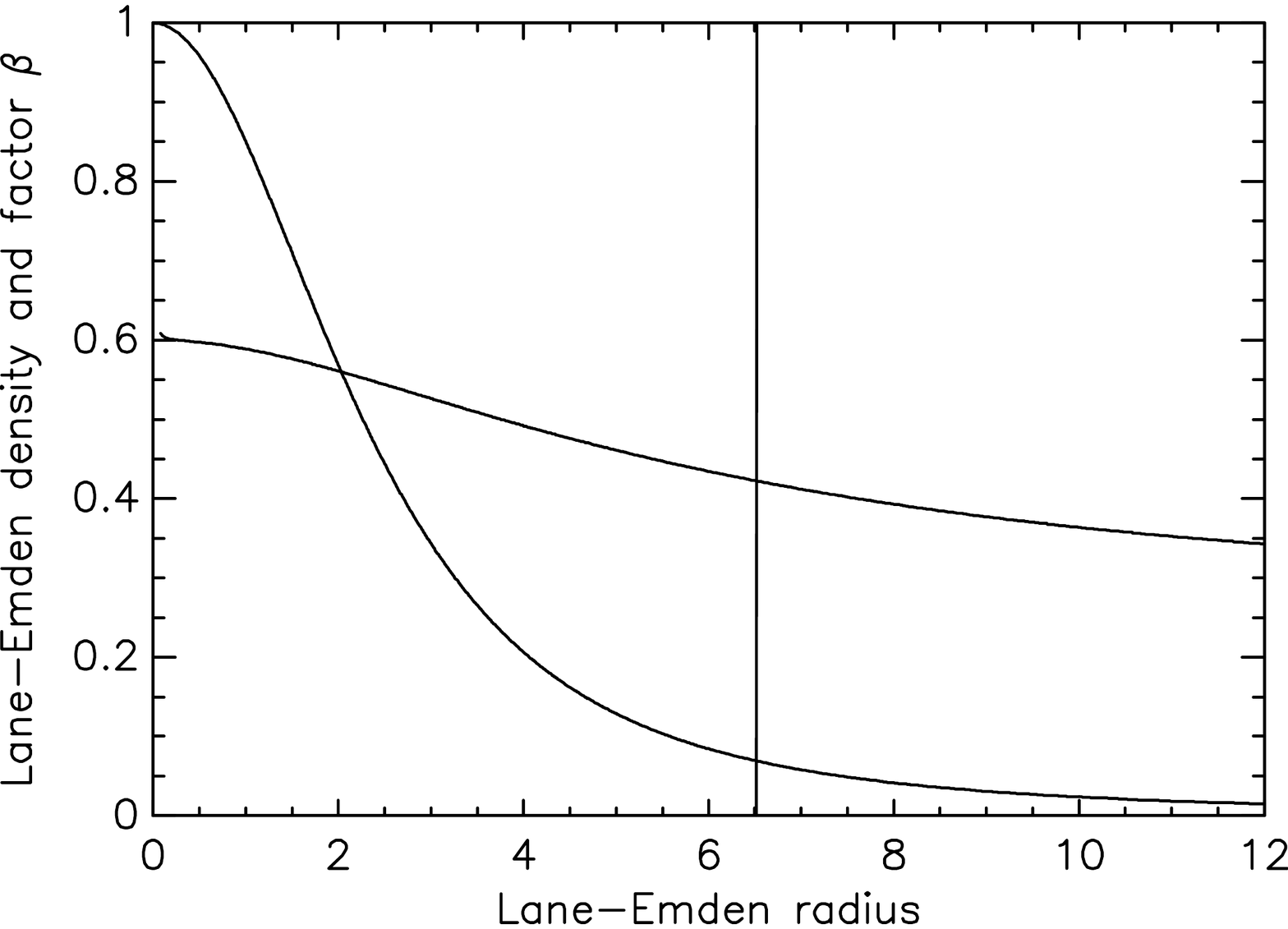}
%\vskip -0.5in
\caption{
The factor for the moment of inertia, $\beta$ as a function of the non-dimensional radius in the Lane-Emden equation. 
Also shown is the non-dimensional hydrostatic density profile described by this equation and normalized to the central density.
In this case $\beta$ varies as a function of radius between 0.6 and 0.4
if the cloud is truncated at the critical radius of a Bonnor-Ebert sphere shown by the vertical line. 
}
\label{beta-plot}
\end{figure}

The factor $\beta$ is defined by equation \ref{betadef}.
\begin{equation}
\beta \equiv \frac{I} {MR^2} = \frac {\int \rho r^4 dr} {R^2 \int \rho r^2 dr}
\end{equation}
For a power-law density profile $\rho \propto r^{-p}$,
\begin{equation}
\beta = \frac{3-p} {5-p}.
\end{equation}
So $\beta= 0.6$ if $p=0$ (constant density) and $\beta=1/3$ if $p=2$. The latter is characteristic of isothermal clouds in self-gravitating equilibrium.
Figure \ref{beta-plot} plots $\beta$ versus radius for a hydrostatic profile. For example, $\beta = 0.42$ for a Bonner-Ebert sphere and 0.6
for a sphere of uniform density
Values of $\beta$ between 0.3 and 0.6 cover most clouds of interest for
our calculations.
%/Users/dog/papers/field/Sept-18/LE/be.pro

\section{Free-fall time}\label{APPENDIXB}

The free-fall time in our non-dimensional units is calculated from equation \ref{ndvirial} deleting the term for the kinetic energy,
\begin{equation}\label{ff1}
\frac {d^2 r^2}{d\tau^2} = -\frac{6}{\beta r} .
\end{equation}
With 
\begin{equation}\label{drdt}
v = \frac{dr}{d\tau}
\end{equation}
equation \ref{ff1} is,
\begin{equation}
2 r \frac{dv}{d\tau} +  v^2 = -\frac{6}{\beta r}.
\end{equation}
Noting that,
\begin{equation}
\frac{d}{dr} (r^2v^2) = 2r^2 \frac{dv}{d\tau} + v^2 2r
\end{equation}
we can separate the variables, 
\begin{equation}
\frac{d}{dr} (r^2v^2) = -\frac{6}{\beta}
\end{equation}
and integrate between the outer radius, $r_0=1$ and $v_0=0$, and $r$, to obtain,
\begin{equation}
v = \frac{1}{r} \bigg( \frac{6}{\beta} \bigg)^{1/2} (1-r)^{1/2} .
\end{equation}
Use equation \ref{drdt} and integrate again,
to get the free-fall time,
\begin{equation}
\tau_F = \frac{4}{3} \bigg( \frac{\beta}{6} \bigg)^{1/2}
\end{equation}
or $\tau_F \approx \frac{1}{3}$ if $\beta = 0.4$.

%%%%%%%%%%%%%%%%%%%%%%%%%%%%%%%%%%%%%%%%%%%%%%%%%%

% Don't change these lines
\bsp	% typesetting comment
\label{lastpage}
\end{document}